\documentclass{webofc}

\usepackage[varg]{txfonts}   
\usepackage{hyperref}
\usepackage{url}
\hypersetup{colorlinks=true,citecolor=blue,urlcolor=blue,linkcolor=blue}
\begin{document}
\title{Magnetic Field Effects on Hadron Yields and Fluctuations}
%
%
%

\author{
\firstname{Volodymyr} \lastname{Vovchenko}\inst{1}\fnsep\thanks{\email{vvovchenko@uh.edu}}
}

\institute{Physics Department, University of Houston, Box 351550, Houston, TX 77204, USA}

\abstract{We explore the impact of an external magnetic field on hadron yields and fluctuations in a thermalized system within the hadron resonance gas (HRG) model using an expanded \texttt{Thermal-FIST} package.
The magnetic field 
sizably enhances the final proton-to-pion (\( \text{p}/\pi \)) ratio due to $\Delta(1232)^{++}$ decay feeddown 
indicating that this ratio may serve as a potential magnetometer for freeze-out conditions. 
At the same time, magnetic field does not generate additional dynamical fluctuations of hadron numbers. 
This suggests that fluctuations in heavy-ion collisions provide limited additional information about the magnetic field beyond what is already inferred from mean multiplicities.
}
\maketitle
\section{Introduction}
\label{sec:intro}

The properties of hot QCD matter change under the presence of an external magnetic field $eB$~\cite{Andersen:2014xxa,Miransky:2015ava,Endrodi:2024cqn}.
Lattice QCD simulations indicate inverse magnetic catalysis of the chiral transition temperature at and a possibility of a critical point in temperature-magnetic field plane~\cite{Bali:2011qj,Bali:2012zg,Endrodi:2015oba,DElia:2021yvk}.
Recent lattice QCD simulations evaluate the effect of magnetic field on conserved charge susceptibilities~\cite{Ding:2021cwv}, with a sizable influence of baryon-charge correlator.
It has been argued that this quantity may serve as a magnetometer in heavy-ion collisions~\cite{Ding:2023bft}, where significant magnetic fields are generated in the initial state, and may potentially survive to later stages of the collision~\cite{Kharzeev:2007jp,Tuchin:2013ie}.
Here, we re-evaluate the possible effect of the magnetic field on hadron yields and fluctuations by employing the hadron resonance model.

\section{Hadron resonance gas in external magnetic field}
\label{sec:HRG}

The hadron resonance gas (HRG) model offers a well-established theoretical framework for studying the thermodynamics of hadronic matter. 
By incorporating external magnetic field into this model, one can investigate how it affects hadronic observables, such as yield ratios and conserved charge fluctuations.
The HRG model in an external magnetic field $eB$ has been formulated in Ref.~\cite{Endrodi:2013cs}, and this formulation is utilized here. 
The presence of magnetic field leads to the modification of the energy levels and discretization of transverse momenta for charged hadrons. 
It also leads to vacuum contribution to thermodynamics, which is, however, independent of chemical potentials and thus omitted in the present work.
The partial pressure of charged hadron species $i$ reads
\begin{align}
\label{eq:pc}
    p_c^i 
    & = \frac{|q_i| B T}{2 \pi^2} \sum_{s_z = -s}^s \sum_{l=0}^{\infty} \varepsilon_c \sum_{k=1}^{\infty} (-\eta_i)^{k+1} \frac{e^{k\mu_i/T}}{k} K_1(k\varepsilon_c/T).
\end{align}
Here $\varepsilon_c = \sqrt{m^2 + 2 |q| B (l + 1/2 - s_z)}$, $s_z$ is the spin of hadron species $i$, $m_i$ is its mass, $\mu_i = b_i \mu_B + q_i \mu_Q + s_i \mu_S$ is its chemical potential, and $\eta_i = \pm 1$ corresponds to Fermi~(Bose) statistics.
The hadron number density is calculated as $n_i = (\partial p_c^i / \partial \mu_i)$ while the susceptibilities of conserved charges are given by $\chi^{BQS}_{lmn} = \frac{\partial^{l+m+n}(p/T^4)}{\partial(\mu_{B}/T)^{l} \, \partial(\mu_{Q}/T)^{m} \, \partial(\mu_{S}/T)^{n}}$.

The effect of the magnetic field on HRG thermodynamics at finite temperature has been implemented in the open-source \texttt{Thermal-FIST} package~\cite{Vovchenko:2019pjl} since version 1.5~(see \cite{Vovchenko:2024wbg} for details).
The present implementation is based on Ref.~\cite{Endrodi:2013cs} and does not include additional expected magnetic field effects such as modification of hadron masses~(including neutral hadrons~\cite{Coppola:2019uyr}) and decay rates~\cite{Marczenko:2024kko}.

\section{Hadron yields}
\label{sec:yields}

\begin{figure}[t]
\centering
\includegraphics[width=.58\textwidth,clip]{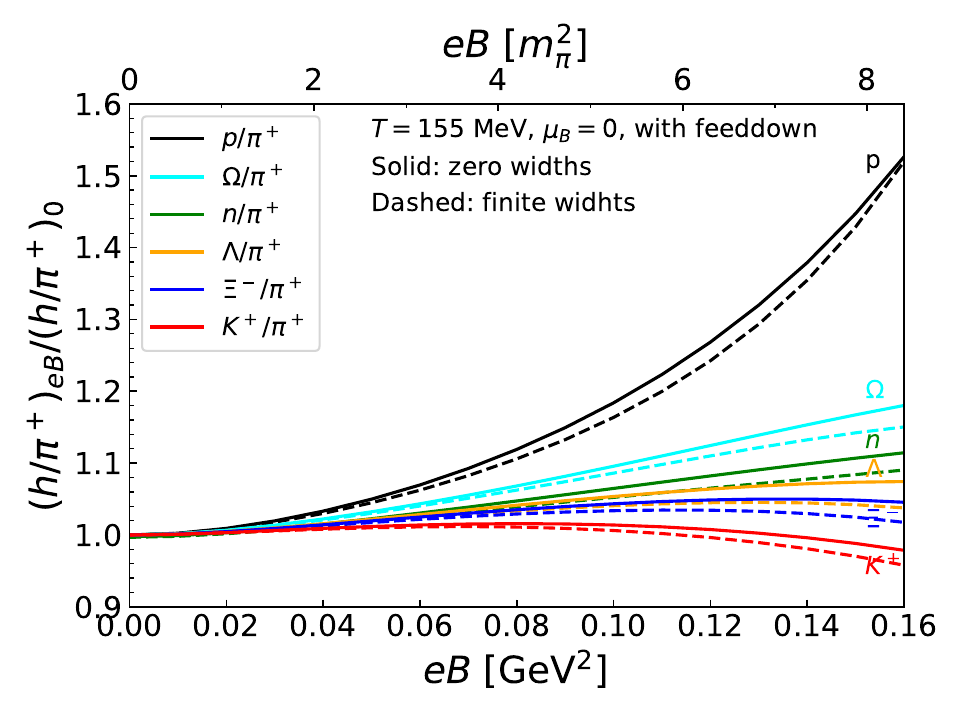}
\includegraphics[width=.37\textwidth,clip]{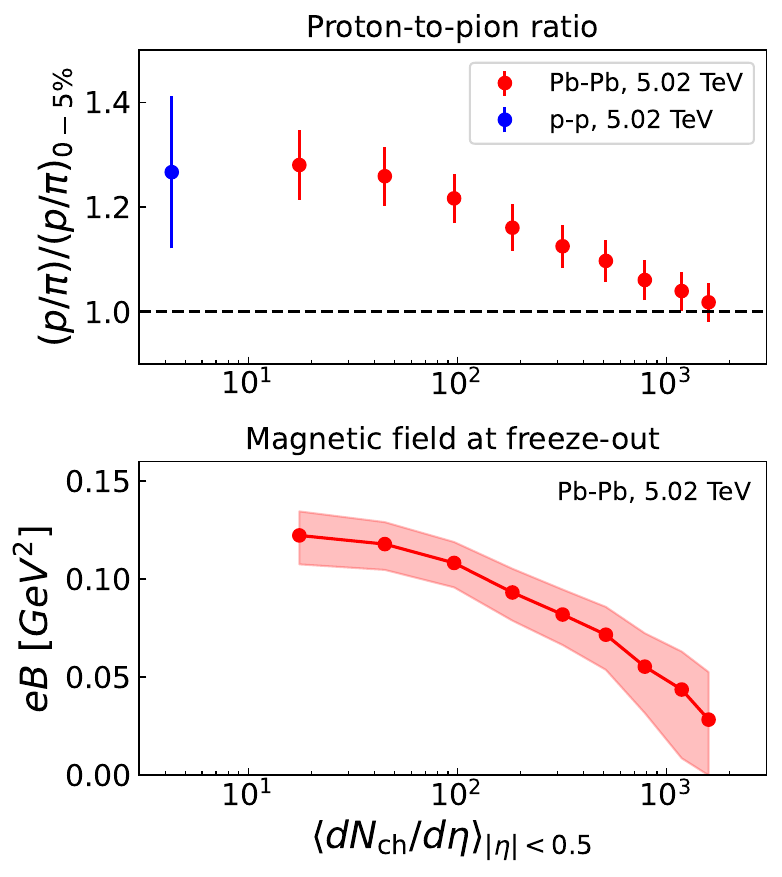}
\caption{
\emph{Left:} Dependence of various final hadron-to-pion yield ratios on the external magnetic field $eB$ in HRG model at temperature $T = 155$~MeV and vanishing chemical potentials, with~(solid) and without~(dashed) finite resonance widths. 
The ratios are normalized by their values at $eB = 0$.
\emph{Right:} Centrality dependence of the proton-to-pion ratio in Pb-Pb collisions at the LHC~\cite{ALICE:2019hno} and the extracted magnetic field necessary to describe the experimental data.
}
\label{fig:ratios}       
\end{figure}

{\bf Enhancement of the densities.} 
Primordial hadron number densities are computed as chemical potential derivatives of the pressure in Eq.~\eqref{eq:pc}, which includes the effect of the magnetic field.
In heavy-ion collisions, the final yields contain feeddown contributions from strong and electromagnetic decays.
These are computed here assuming that magnetic field leaves decay branching ratios unaffected.
Figure~\ref{fig:ratios} depicts the dependence of various hadron-to-pion ratios computed at LHC conditions~($T = 155$~MeV, $\mu_B = 0$) as a function of the magnetic field $eB$.
The calculations incorporate feeddown from strong and electromagnetic decays, with~(dashed) or without~(solid) finite resonance widths through energy-dependent Breit-Wigner prescription~\cite{Vovchenko:2018fmh}.
The effect of finite widths is mild.
The main observation is the strong enhancement of the $p/\pi$ ratio.
This can be attributed to the feeddown from $\Delta(1232)^{++}$, which is strongly enhanced by the magnetic field due to its large spin~(3/2) and charge~($q=2$).
Treatment of high-spin states should be improved, however, as the analysis of vacuum contributions to the pressure within the present framework indicates that $S = 3/2$ (and above) states may make the system unstable~\cite{Endrodi:2013cs}.

{\bf Proton-to-pion ratio as a magnetometer.} 
Taking the calculated strong effect of magnetic field on the p/$\pi$ ratio at face value, one can expect the magnetic field to induce sizable centrality dependence of this ratio,  with larger enhancements observed in peripheral collisions where the magnetic field is strongest.
Attributing the observed enhancement of the p/$\pi$ ratio as a function of centrality reported by the ALICE Collaboration for 5.02 TeV Pb-Pb collisions~\cite{ALICE:2019hno} entirely to the change of magnetic field, one can estimate the strength of the magnetic field at chemical freeze-out.
This is shown in right panel of Fig.~\ref{fig:ratios}, indicating the maximum magnetic field in peripheral collisions of $eB \simeq (0.13 \pm 0.01)\text{GeV}^2 \simeq (6.6 \pm 0.6) m_\pi^2$.

However, several important caveats need to be considered in this interpretation. The assumption of a constant magnetic field throughout the fireball at freeze-out may not be entirely accurate, as large event-by-event fluctuations of the magnetic field are expected~\cite{Bzdak:2011yy}. Perhaps even more importantly, other effects, such as baryon annihilation in the hadronic phase, can also induce centrality dependence of the p/$\pi$ ratio~\cite{Vovchenko:2022xil}, and must be carefully disentangled from the magnetic field's influence.

{\bf Isospin-symmetry breaking and neutron yields.} 
In addition to the p/$\pi$ ratio, other hadron yield ratios are also affected by the magnetic field. 
Neutron yields are particularly interesting. 
In the isospin-symmetric scenario, the yields of neutrons and protons are expected to be identical.
On the other hand, magnetic field does induce the breaking of the isospin symmetry.
This is reflected in Fig.~\ref{fig:ratios} by a much weaker enhancement of the n/$\pi$ ratio compared to p$/\pi$ due to magnetic field.
Thus, the corresponding measurements of n/p ratio and its possible deviations from unity as a function of centrality provide a more unambiguous probe of the magnetic field compared to the p/$\pi$ ratio, motivating its challenging experimental measurement in the future.

\section{Fluctuations and correlations}
\label{sec:flucs}

\begin{figure}[t]
\centering
\includegraphics[width=.33\textwidth,clip]{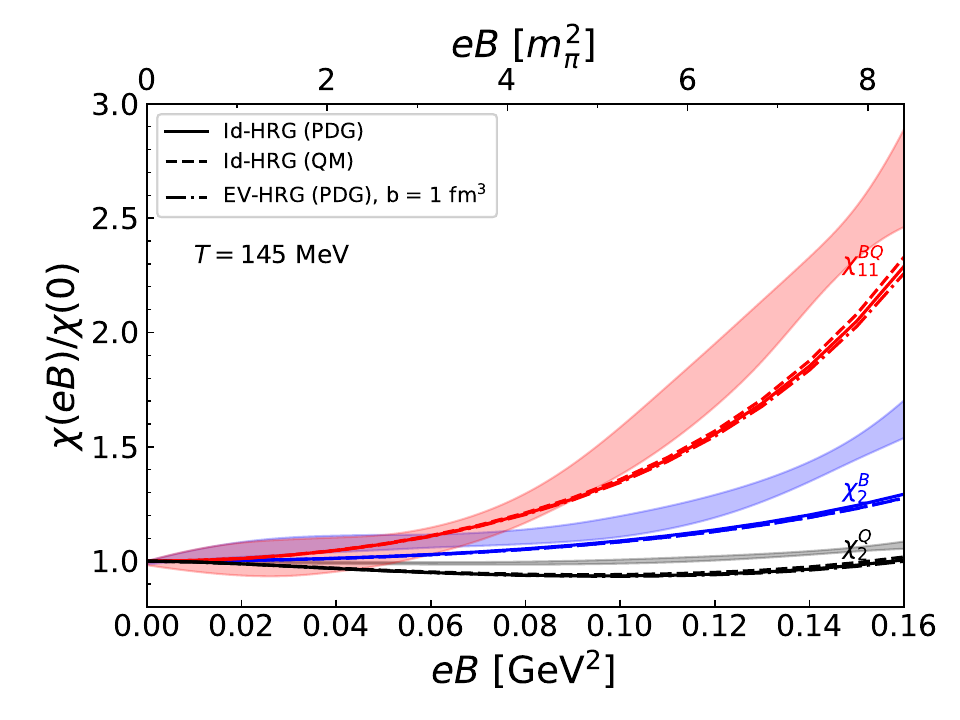}
\includegraphics[width=.29\textwidth,clip]{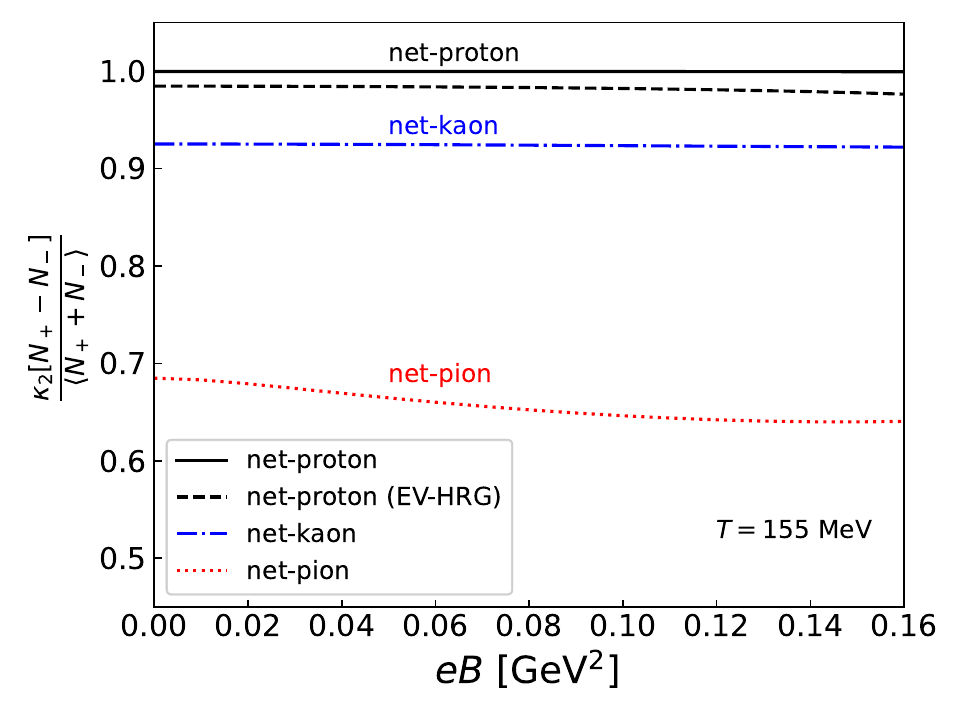}
\includegraphics[width=.33\textwidth,clip]{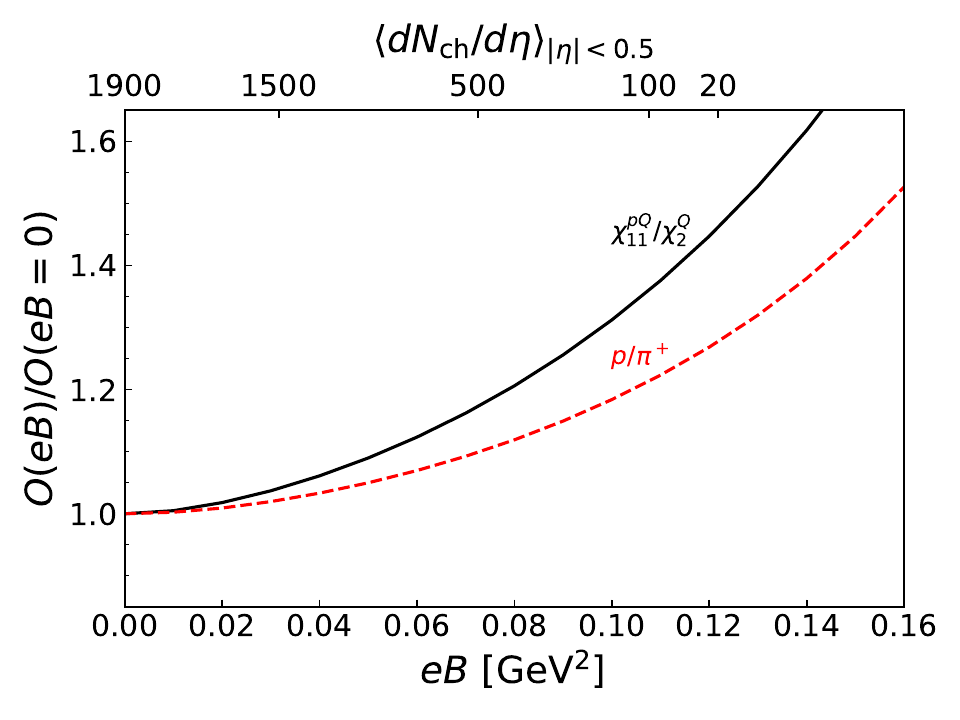}
\caption{
\emph{Left:} Dependence of second-order susceptibilities of conserved charges on $eB$ within different versions of the HRG model and comparison with lattice QCD at $T  = 145$~MeV.
\emph{Middle:} Normalized variance of final net-proton, net-kaon, and net-pion numbers as a function of $eB$ at $T = 155$~MeV.
\emph{Right:} Normalized proton-charge correlator, $\chi_{11}^{pQ} / \chi_2^Q$, and the p/$\pi$ ratio as a function of the $eB$ at $T = 155$~MeV.
}
\label{fig:flucs}       
\end{figure}

Comparing HRG model predictions for susceptibilities of conserved charges with lattice QCD allows one to probe the validity range of HRG. 
Figure~\ref{fig:flucs} depicts the magnetic field dependence of $\chi_{11}^{BQ}$, $\chi_2^B$, and $\chi_2^Q$ at a temperature $T = 145$~MeV calculated using the HRG model with different hadron spectra and baryon excluded volume effects, and compared to lattice QCD simulations~\cite{Ding:2023bft}.
The comparison indicates qualitative agreement of the HRG model predictions with lattice QCD, which both predict the enhancement of $\chi$'s with $eB$, but also the need for quantitative improvement of the model, especially at $eB \gtrsim 0.08$~GeV$^2$. 

HRG model calculations indicate that hadron number densities entirely drive the enhancement of susceptibilities with the magnetic field. 
When the corresponding fluctuations are normalized by the means -- the Skellam baseline -- the effect of magnetic field is essentially canceled out.
This is illustrated by the middle panel in Fig.~\ref{fig:flucs}, where the normalized variance $\kappa_2[N_+ - N_-]/\langle N_+ + N_- \rangle$ of net-proton (solid black), net-kaon~(dash-dotted blue), and net-pion~(dotted red) numbers is shown as a function of $eB$.
These quantities exhibit essentially flat dependence on the magnetic field, and the observed deviations from unity are driven by correlations from resonance decays rather than the magnetic field.

Finally, the normalized proton-charge correlator, $\chi_{11}^{pQ} / \chi_2^Q$ can be considered as a proxy for the ratio $\chi_{11}^{BQ} / \chi_2^Q$ that is enhanced by the magnetic field.
This ratio is shown in Fig.~\ref{fig:flucs} as a function of $eB$ together with the p/$\pi$ ratio.
One can see that $\chi_{11}^{pQ} / \chi_2^Q$ is indeed enhanced by the magnetic field, and it is primarily driven by the enhancement of the p/$\pi$ ratio, given that $\chi_{11}^{pQ}$ and $\chi_{11}^{pQ}$ are dominated by the proton and pion self-correlation terms, respectively.
The relative enhancement of $\chi_{11}^{pQ} / \chi_2^Q$ is stronger than that of the $p/\pi$ ratio, which can be attributed to the $\Delta^{++} \to p + \pi^+$ decays.
Thus, the relevance of the magnetic field could potentially be probed by precision measurements of the difference between $\chi_{11}^{pQ} / \chi_2^Q$ and $p/\pi^+$ ratios.

\section{Summary and Outlook}
\label{sec-5}
We have investigated the effects of magnetic fields on hadron yields and fluctuations in the hadron resonance gas model. Our results indicate that magnetic fields significantly influence select hadron yields, particularly by enhancing the final proton-to-pion ratio via the decay feeddown of $\Delta(1232)^{++}$, which can potentially serve as a magnetometer in heavy-ion collisions. 
The neutron-to-proton ratio can be a clearer probe of the magnetic field due to its isospin symmetry-breaking nature, though it is more challenging to measure experimentally.
While magnetic fields have a substantial impact on hadron yields, their effect on normalized fluctuations of hadron yields is minimal.
This suggests that hadron yield ratios may serve as better probes of the magnetic field in heavy-ion collisions than fluctuations.

\end{document}